\documentclass[12 pt, preprint]{aastex}
\begin{document}

\title{New solar opacities, abundances, helioseismology, and neutrino fluxes }
\author{John N. Bahcall and Aldo  M. Serenelli}
 \affil{Institute for Advanced Study, Einstein Drive,
Princeton, NJ
  08540}
\and
\author{Sarbani Basu}
\affil{Department of Astronomy, Yale University, New Haven, CT
06520-8101}

\begin{abstract}
We construct solar models with the newly calculated radiative
opacities from the Opacity Project (OP) and recently determined
(lower) heavy element abundances. We compare results from the new
models with predictions of a series of models that use OPAL
radiative opacities, older determinations of the surface heavy
element abundances, and refinements of nuclear reaction rates. For
all the variations we consider, solar models that are constructed
with the newer and lower heavy element abundances advocated by
Asplund et al. (2005)\nocite{asplundgrevessesauval2005} disagree
by much more than the estimated measuring errors with
helioseismological determinations of the depth of the solar
convective zone, the surface helium composition, the internal
sound speeds, and the density profile. Using the new OP radiative
opacities, the ratio of the $^8$B neutrino flux calculated with
the older and larger heavy element abundances (or with the newer
and lower  heavy element abundances) to the total neutrino flux
measured by the Sudbury Neutrino Observatory is 1.09 (0.87) with a
9\% experimental uncertainty and a 16\% theoretical uncertainty,
$1\sigma$ errors.
\end{abstract}

\keywords{Sun: abundances, atomic processes, neutrinos, nuclear
reactions, Sun: interior}

\maketitle

Recent, refined determinations of the surface heavy element abundances
of the Sun have led to lower than previously believed heavy element
abundances~(see Asplund et al. 2005 and references
therein).\nocite{asplundgrevessesauval2005} A number of authors have
pointed out that these lower heavy element abundances lead to solar
models that conflict with different aspects of helioseismological
measurements (e.g., Bahcall \& Pinsonneault 2004\nocite{BP04}, Basu
 \& Antia 2004\nocite{basu04}, Bahcall et al. 2005)\nocite{BBPS05}. If the radiative opacity in the
temperature range of $2\times 10^6$K to $4.5 \times 10^6$K were to be
increased by of order 10\% relative to the standard OPAL opacity
(Iglesias \& Rogers 1996 ),\nocite{opalopacity96} then the discrepancy
between new abundances and helioseismology could be resolved (Bahcall
et al. 2005,\nocite{BBPS05} see also Basu
 \& Antia 2004)\nocite{basu04}.

The Opacity Project (OP) has recently performed more precise and
more physically complete calculations of the radiative opacities
with the goal of determining if these new calculations could
eliminate the discrepancy between helioseismology and solar
modeling that uses the new (lower) heavy element abundances (see
Badnell et al. 2004\nocite{badnelletal2004}; Seaton \& Badnell
2004\nocite{seaton}; Seaton 2004\nocite{seatononly}). The Opacity
Project refinements result in only a small increase (less than
2.5\% everywhere of interest) relative to the OPAL opacity.

In this paper, we present a series of precise solar models that were
calculated using the new OP opacities as well as with the familiar
OPAL opacities. We also present models that were constructed with the
recently-determined heavy element abundances (Asplund et
al. 2005)\nocite{asplundgrevessesauval2005} as well as with the
previously standard abundances (Grevesse \& Sauval
1998).\nocite{oldcomp} In addition, we introduce refinements in the
nuclear physics used in the solar models.  We compare the results of
each of our series of solar models with helioseismological and
neutrino observations of the Sun. As a side-product of this
investigation, we determine the remarkable precision with which two
very different stellar evolution codes reproduce the same solar model
parameters.

\begin{figure} [t!]
\begin{center}
\includegraphics[width=9cm]{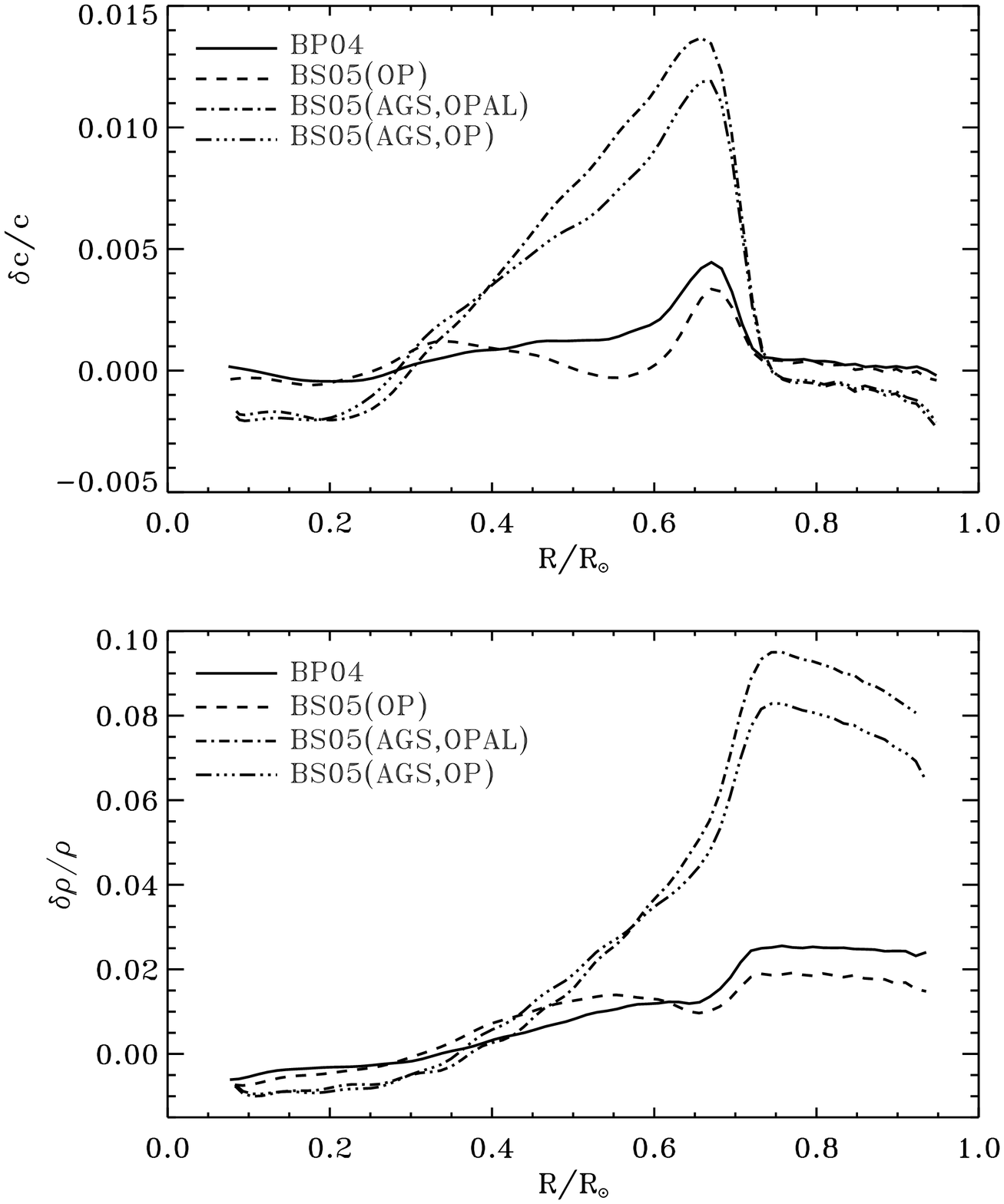}
\end{center}
\caption{Relative  sound speed  differences,  $\delta c/c=(c_\odot
-  c_{\rm
    model})/c_{\rm  model}$,  and relative densities, $\delta
    \rho/\rho$,
    between   solar  models  and  helioseismological
    results from MDI data.
\label{fig:velocitydiffs}}
\end{figure}

 Table~\ref{tab:modelparameters} gives the
principal characteristics of seven precise solar models that we
use in this paper to investigate the helioseismological and
neutrino flux implications of the recent redeterminations of heavy
element abundances and of radiative opacities.
Table~\ref{tab:neutrinofluxes} presents the neutrino fluxes
calculated for each of the seven solar models represented in
Table~\ref{tab:modelparameters}.  At the end of the paper, we
summarize in Figure~\ref{fig:velocitydiffs} and the related
discussion the comparison between the helioseismologically
determined sound speeds and densities and the predictions of the
various solar models. We begin by  describing the differences
between the various solar models and by commenting on how these
differences affect the calculated properties of the models,
including helioseismological parameters and neutrino fluxes.

The model BP04(Yale) was calculated by Bahcall \& Pinsonneault
(2004)\nocite{BP04} and is their preferred standard solar model.
BP04(Yale) uses the Grevesse \& Sauval (1998)\nocite{oldcomp}
solar abundances and the best other input data available when the
model was constructed. The model was constructed as described in
Bahcall \& Pinsonneault (1992)\nocite{BP92} and Bahcall \& Ulrich
(1988)\nocite{bahcallulrich1988} and uses the Yale-Ohio
State-Princeton stellar evolution code (Pinsonneault et al.
1989;\nocite{pinsonneault89} Bahcall \& Pinsonneault
1992\nocite{BP92}, 1995)\nocite{BP95} as modified by iterations of
the Bahcall-Ulrich nuclear energy generation subroutine. The model
BP04(Garching) was derived using the Garching Stellar Evolution
code (see, e.g., Schlattl, Weiss, \& Ludwig 1997\nocite{garching}
and Schlattl 2002\nocite{schlattl02} for details of the code)
using the same procedures and input data as the BP04(Yale) solar
model.

The first two rows of Table~\ref{tab:modelparameters} and
Table~\ref{tab:neutrinofluxes} show that the principal
characteristics of solar models are independent, to practical
accuracy, of the evolutionary code used for their calculation. For
example, the initial helium abundance is the same in the BP04
(Yale) and BP04 (Garching) models to an accuracy of $\pm 0.04$\%
and the depth of the convective zone is the same to $\pm 0.01$\%.
In a more stringent test, the $^7$Be,$^8$B, $^{17}$F , and pep
neutrino fluxes in the two models agree to $\pm 0.4$\% or better
and the p-p, hep,$^{13}$N, and $^{15}$O neutrino fluxes to better
than $\pm 0.1$\%. This important result, which demonstrates that
two different stellar evolution codes yield the the same answers
to high precision, shows that we have to take seriously
discrepancies between solar model predictions and observations
even when the discrepancies are very small.

The  small differences between  the  BS04 and the
\hbox{BP04(Garching)} solar models can  be summarized   as
follows. First, in the BS04
 model, individual metals diffuse at the different velocities
implied by  the Thoul, Bahcall, \& Loeb (1994)\nocite{thoul4}
analysis, whereas in the BP04 calculation all the metals are
assumed to diffuse at the same velocity (usually taken to be that
of the iron). The changes in abundances induced by using
individual velocities are very small, parts per thousand. Second,
in the BS04 model, the increase in metallicity caused by the
burning of $^{12}$C that is out-of-CN-equilibrium into $^{14}$N is
accounted for in the evaluation of the radiative opacities. Two
protons are included together with $^{12}$C in the conversion to
$^{14}$N. In the Garching code, this increase in $Z$ is taken into
account whereas in the Yale code the change in composition is
added to the helium abundance. Third, because $^{17}$O burns
slowly at the solar center temperatures, the $^{17}$O abundance is
not assumed to be in equilibrium in the BS04 model and is
essentially unmodified after it is produced by
$^{16}$O(p,$\gamma$)$^{17}$F and the beta-decay of $^{17}$F. In
the Yale code, the reaction $^{17}$O(p,$\alpha$)$^{14}$N is
assumed to occur very fast due to a resonant reaction.

The refinements in physics between the BS04 and the
\hbox{BP04(Garching)} models do not change significantly the
computed astronomical characteristics that are summarized in
Table~\ref{tab:modelparameters}. For example, the initial helium
abundances inferred from the BP04(Yale), BP04(Garching), and BS04
models all agree to about $\pm 0.1$\% and the other astronomical
characteristics are, in nearly all cases, the same in all three
models to comparable or better accuracy. The neutrino fluxes are
practically the same in all three models, with the most important
change, $\pm 1$\%, occurring for the $^8$B neutrino flux.

In what follows, we will discuss solar models constructed with the
Garching code and will denote the different models by BS05 (plus
specifications). Each successive improvement will be incorporated
in all subsequent models except where noted otherwise.

The model BS05($^{14}$N) is the same as the model BS04 except that
in the newer model we use the recently measured value of $S_{1,14}
= 1.7 \pm 0.2$ keV b for the low energy cross section factor of
the $^{14}$N(p,$\gamma$)$^{15}$O fusion reaction (Formicola et al.
2004).\nocite{14N} Again, this improvement makes no practical
change in the traditional astronomical characteristics of the
model that are shown in Table~\ref{tab:modelparameters}. However,
BS05($^{14}$N) has $^{13}$N and $^{15}$O solar neutrino fluxes
that are almost a factor of two lower than the corresponding
fluxes obtained from the BS04 solar model. The  CNO contribution
to the solar luminosity is also reduced compared to BS04,
BP04(Garching), and BP04(Yale). The latter models have a CNO
contribution of 1.55\% to the solar luminosity while for
BS05($^{14}$N) the CNO contribution is only 0.8\%.

The next two solar models are the first in the series to use OP
opacities. BS05(OP) and BS05(AGS, OP) differ in that BS05(AGS, OP)
uses the heavy element abundance taken from Asplund et al.
(2005).\nocite{asplundgrevessesauval2005} Like all the proceeding
models, BS05(OP) uses Grevesse \& Sauval (1998)\nocite{oldcomp}
abundances. Comparing BS05(OP) with BS05($^{14}$N), we see that
the new OP opacities do not change significantly the neutrino
fluxes nor other principal model characteristics.

The lower heavy element abundances used in BS05(AGS,OP)cause the
computed depth of the convective zone to be too shallow and the
surface helium abundance to be unacceptably low, as compared with
the helioseismologically measured values. The depth of the solar
convective zone and the helium surface abundance have recently
been redetermined by Basu \& Antia (2004).\nocite{basu04} using
the best-available helioseismological data. Comparing the values
calculated using BS05(AGS,OP) with the measured values (given in
parentheses), we have

\begin{eqnarray}
\frac{R_{\rm CZ}}{R_\odot} & = &0.728 (0.713 \pm 0.001,\, {\rm
exp.}); \\ ~ Y_{\rm surf} & = & 0.229 (0.249 \pm 0.003,\, {\rm
exp.})\, . \label{eq:czhelium}
\end{eqnarray}
For BS05(AGS,OP), the disagreements between helioseismological
measurements and the computed values of $R_{\rm CZ}$ and $Y_{\rm
surf}$ are many times the quoted errors. By contrast, all of the
models in Table~\ref{tab:modelparameters} that use the Grevesse \&
Sauval (1998)\nocite{oldcomp} abundances [BP04(Yale),
BP04(Garching), and BS04, BS05($^{14}$N), BS05(OP)] have values
for these parameters, $R_{\rm CZ}\sim 0.715$ and $Y_{\rm surf}
\sim 0.244$, that are in much better agreement with
helioseismological measurements.

Similar results are obtained with models that use OPAL opacities
(see row labeled BS05(AGS, OPAL) in
Table~\ref{tab:modelparameters}). Solar models constructed with
the AGS05 composition disagree with the helioseismological
measurements of $R_{\rm CZ}$ and  $Y_{\rm surf}$, independent of
whether on uses OPAL or OP radiative opacities.

Figure~\ref{fig:velocitydiffs} shows that, for four representative
models, the sound speeds and densities inferred from solar models
that use the Grevesse \& Sauval (1998)\nocite{oldcomp} solar
abundances are in excellent agreement with the helioseismological
measurements (Schou et al. 1998)\nocite{sch98} of sound speeds and
densities. Solar models that use the new Asplund, Grevesse, \&
Sauval (2005)\nocite{asplundgrevessesauval2005} abundances are in
disagreement with the helioseismological measurements. For models
that use the Grevesse \& Sauval (1998)\nocite{oldcomp} abundances
and OPAL, the rms difference between the solar model predictions
for sound speeds and densities are, respectively, $0.0015 \pm
0.0001$ and $0.015 \pm 0.002$, where we quote the range that spans
the values for the first four models that appear in
Table~\ref{tab:modelparameters}. The results with OP opacities are
even better: 0.00097 and 0.012, respectively. By contrast, the rms
differences for models that use the AGS05 abundances are larger by
more than a factor of three, $0.0053 \pm 0.0005$ and $0.047 \pm
0.003$, respectively.

\begin{figure} [ht!]
\begin{center}
\includegraphics[angle=270,width=9cm]{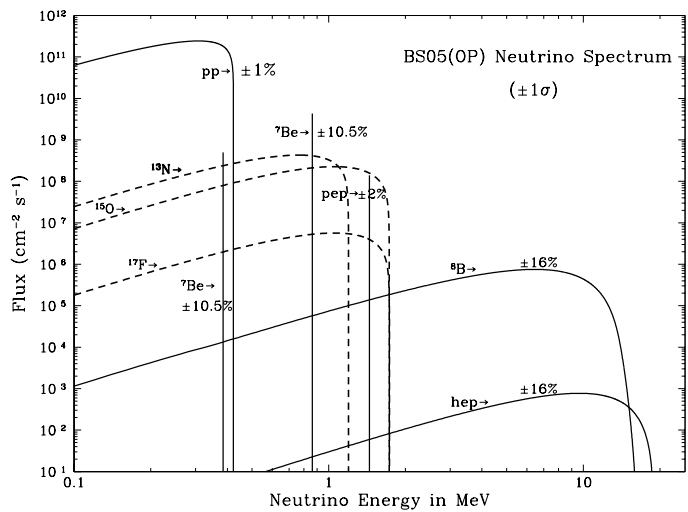}
\end{center}
\caption{Solar neutrino energy spectrum for the solar model
BS05(OP). The uncertainties are taken from Table~8 of Bahcall and
Serenelli (2005). \label{fig:energyspectrum}}
\end{figure}

How do the adopted element abundances and the radiative opacity
affect the predicted solar neutrino fluxes?
Figure~\ref{fig:energyspectrum} shows the solar neutrino energy
spectrum that is calculated using the BS05(OP) solar model, which
may be taken as the currently preferred solar model. The
fractional uncertainties for the neutrino fluxes are given in
Table~8 of Bahcall and Serenelli (2005).\nocite{bs05}

Using OP opacity, the ratio of the $^8$B neutrino flux calculated
with the older (larger) heavy element abundances (or with the
newer (lower) heavy element abundances) to the total $^8$B
neutrino flux measured by the Sudbury Neutrino Observatory (Ahmed
et al. 2004)\nocite{snosalt} is (see
Table~\ref{tab:neutrinofluxes})
\begin{equation}
\frac{{\rm Solar \, Model}~^8{\rm B} ~\nu~ {\rm flux}}{{\rm
Measured}~^8{\rm B}~ \nu~ {\rm flux}} ~=~ 1.09 (0.87) \, ,
\label{eq:b8ratio}
\end{equation}
with a 9\% experimental error (Ahmed et al. 2004)\nocite{snosalt}
and a 16\% theoretical uncertainty (Bahcall and Serenelli 2005),
$1\sigma$ uncertainties. If we adopt OPAL opacities, the
coefficients on the right hand side of equation~(\ref{eq:b8ratio})
become 1.12 (0.88), very similar to the values for OP opacities.
Turck-Chieze et al. (2004) found a 9\% lower $^8$B neutrino flux
for a model similar to  BS05(AGS, OPAL).  Their lower flux is
accounted for by the fact that Turck-Chieze et al. did not use the
recent and more accurate pp cross section calculated by Park et
al. (2003) and Turck-Chieze et al.  did use intermediate screening
for fusion reactions instead of the more accurate approximation of
weak screening (see Bahcall, Brown, Gruzinov, and Sawyer 2002).

Comparing the calculated to the measured (Bahcall,
Gonzalez-Garcia, \& Pe\~na-Garay 2004)\nocite{BP-G04} p-p neutrino
flux, assuming OP opacities, we have
\begin{equation}
\frac{{\rm Solar \, Model}~{\rm p-p} ~\nu~ {\rm flux}}{{\rm
Measured}~{\rm p-p}~ \nu~ {\rm flux}} ~=~ 0.99 (1.00) \, ,
\label{eq:ppratio}
\end{equation}
with a 2\% experimental uncertainty (Bahcall, Gonzalez-Garcia, \&
Pena-Garay 2004)\nocite{BP-G04} and a 1\% theoretical uncertainty
(Bahcall and Serenelli 2005). The agreement is similarly good if
we adopt OPAL opacities. The CNO contribution to the solar
luminosity is only 0.5\% for the models BS05(AGS,OP) and
BS05(AGS,OPAL).

 We conclude that
the agreement between solar model predictions and solar neutrino
measurements is excellent and is not significantly affected by the
choice of heavy element abundances or the radiative opacity.

\acknowledgments J. N. B. and A. M. S. are supported in part by NSF grant
PHY-0070928.

\clearpage

\clearpage

\clearpage

\begin{table*}[!t]
\caption{Characteristics of seven solar models.  The table lists the
principal model characteristics for a series of precise solar models
that are defined in the text.  Here $\alpha_{\rm convec}$ is the usual
convective mixing length parameter, $Y_{\rm i}$ and $Z_i$ are the
initial helium and heavy element abundances by mass, $R_{\rm cz}$ is
the radius at the base of the convective zone, $Y_{\rm surf}$ and
$Z_{\rm surf}$ are the present-day surface abundances of helium and
heavy elements, and $Y_{\rm c}$ and $Z_{\rm c}$ are the present-day
abundances at the center of the Sun. The first five models use the
Grevesse \& Sauval (1998) abundances; the last two
models use the Asplund et al. (2005)
abundances. The first four models use OPAL
opacities. \label{tab:modelparameters} }
\begin{center}
\begin{tabular}{lcccccccc}
\hline \hline
 MODEL & $\alpha_{\rm convec}$ & $Y_i$ & $Z_i$ &
$R_{\rm cz}/{\rm R_\odot}$ & $Y_{\rm surf}$ & $Z_{\rm surf}$ & $Y_{\rm c}$ & $Z_{\rm c}$ \\
\hline
BP04(Yale) & 2.07 & 0.2734 & 0.0188 & 0.7147 & 0.243 & 0.0169 & 0.640 & 0.0198 \\
BP04(Garching) & 2.10 & 0.2736 & 0.0188 & 0.7146 & 0.243 & 0.0170 & 0.641 & 0.0196 \\
BS04 & 2.09 & 0.2742 & 0.0188 & 0.7148 & 0.244 & 0.0169 & 0.641 & 0.0202 \\
BS05($^{14}$N) & 2.09 & 0.2739 & 0.0188 & 0.7153 & 0.244 & 0.0170 & 0.635 & 0.0202 \\
BS05(OP) & 2.11 & 0.2725 & 0.0188 & 0.7138 & 0.243 & 0.0170 & 0.634 & 0.0202\\
BS05(AGS,OP) & 1.98 & 0.2599 & 0.0140 & 0.7280 & 0.229 & 0.0126 & 0.620 & 0.0151\\
BS05(AGS,OPAL) & 1.96 & 0.2614 & 0.0140 & 0.7289 & 0.230 & 0.0125 & 0.622 & 0.0151\\
\hline
\end{tabular}
\end{center}
\end{table*}

\clearpage

\begin{table}[!b]
\caption{Predicted solar neutrino fluxes from
seven solar models. The table presents the predicted fluxes, in
units of $10^{10}(pp)$, $10^{9}({\rm \, ^7Be})$, $10^{8}(pep, {\rm
^{13}N, ^{15}O})$, $10^{6} ({\rm \, ^8B, ^{17}F})$, and
$10^{3}({\rm hep})$ ${\rm cm^{-2}s^{-1}}$ for the same solar
models whose characteristics are summarized in
Table~\ref{tab:modelparameters}. \label{tab:neutrinofluxes} }
\begin{tabular}{lcccccccccc}
\hline \hline
Model & pp & pep & hep & $^7$Be &  $^8$B & $^{13}$N & $^{15}$O & $^{17}$F \\
\hline
BP04(Yale) & 5.94 & 1.40 & 7.88 & 4.86 & 5.79 & 5.71 & 5.03 & 5.91 \\
BP04(Garching) & 5.94 & 1.41 & 7.88 & 4.84 & 5.74 & 5.70 & 4.98 & 5.87 \\
BS04 & 5.94 & 1.40 & 7.86 & 4.88 & 5.87 & 5.62 & 4.90 & 6.01 \\
BS05($^{14}$N) & 5.99 & 1.42 & 7.91 & 4.89 & 5.83 & 3.11 & 2.38 & 5.97 \\
BS05(OP)& 5.99 & 1.42 & 7.93 & 4.84 & 5.69 & 3.07 & 2.33 & 5.84 \\
BS05(AGS,OP) & 6.06 & 1.45 & 8.25 & 4.34 & 4.51 & 2.01 & 1.45 & 3.25 \\
BS05(AGS,OPAL) & 6.05 & 1.45 & 8.23 & 4.38 & 4.59 & 2.03 & 1.47 & 3.31 \\
\hline
\end{tabular}
\end{table}

\end{document}